\documentclass[prl,showkeys,showpacs,amsmath,amssymb,twocolumn,superscriptaddress]{revtex4}

\usepackage{graphicx}
\usepackage{graphics}
\usepackage{dcolumn}
\usepackage{bm}
\usepackage{amsmath}

\begin{document}

\preprint{}
\title{Understanding of Ultra-Cold Neutrons Production in Solid Deuterium.}

\author{A. Frei}
\affiliation{Technische Universit\"at M\"unchen,Physik Department,
 James Franck Str., D-85747 Garching, Germany}

\author{E. Gutsmiedl$^{*}$}
\affiliation{Technische Universit\"at M\"unchen,Physik Department,
 James Franck Str., D-85747 Garching, Germany}

\author{C. Morkel}
\affiliation{Technische Universit\"at M\"unchen,Physik Department,
 James Franck Str., D-85747 Garching, Germany}

\author{A.R. M\"uller}
\affiliation{Technische Universit\"at M\"unchen,Physik Department,
 James Franck Str., D-85747 Garching, Germany}

\author{S. Paul}
\affiliation{Technische Universit\"at M\"unchen,Physik Department,
 James Franck Str., D-85747 Garching, Germany}

\author{S. Rols }
\affiliation{Institut Laue Langevin, 156X, F-38042 Grenoble CEDEX,
France}

\author{H. Schober }
\affiliation{Institut Laue Langevin, 156X, F-38042 Grenoble CEDEX,
France}


\author{T. Unruh}
\affiliation{Technische Universit\"at
M\"unchen,Forschungsneutronenquelle Heinz Maier-Leibnitz (FRM II),
Lichtenbergstr. 1, D-85747 Garching, Germany}

\date{\today}

\begin{abstract}
Our recent neutron scattering measurements of phonons and other
quasi-particle excitations in solid deuterium (sD$_2$) and the
extraction of the density of states for phonons and rotational
transitions in sD$_2$ have led us to a new understanding of the
production of ultra-cold neutrons (UCN) in sD$_2$. The UCN
production rate reaches a maximum at an equivalent neutron
temperature of  $T_n$=40~K for a neutron flux with Maxwellian
energy distribution. The cross section for UCN production in
sD$_2$ has been determined by using the density of states $G_1(E)$
in combination with the incoherent approximation as well as by a
direct calibration of our measured neutron cross sections with the
known cross section of the $J=1 \mapsto 0$ rotational transition
in deuterium. Using this cross section we deduced the production
rate of UCN in sD$_2$ which agrees quite well with direct
measurements of this energy averaged UCN production cross section.
\end{abstract}

\pacs{28.20.Cz, 63.20.kk}

\noindent
$^*$Corresponding author; email: egutsmie@e18.physik.tu-muenchen.de\\

\maketitle

Ultra-cold neutrons (UCN) are slow enough to be confined
\cite{Golub1} in traps, which can be formed by materials with a
high Fermi potential (up to~300 neV) or by a magnetic field (60
neV/T). UCN can be observed for several tens of minutes in these
traps, and are excellent tools for high precision measurements, as
 the life time  of the neutron itself \cite{Arz,Ser-1}, and the
search for a possible electric dipole moment of the neutron
\cite{Bak} (current upper limit 2.9$\cdot$10$^{-26}$ e$\cdot$ cm).\\
Powerful UCN sources are needed for the here mentioned
experiments, and different groups \cite{Frei-1,LANL,PSI-1,Ser-2}
are working on the development of strong sources, based on sD$_2$
as an converter for down-scattering of thermal or sub-thermal
neutrons into the UCN energy region (typical $E<$300~neV). A
converter based on sD$_2$ should be operated at temperatures below
T$<$10~K in order to avoid subsequent upscattering of UCN by
phonons within solid deuterium \cite{Liu}. The understanding of
the downscattering of neutrons into the UCN region crucially
relies on a detailed knowledge of the energy loss of thermal or
subthermal neutrons in the sD$_2$ converter material. One major
 reaction channel is based on phonons in the solid deuterium crystal, which are
excited by the neutrons.

We have recently measured the phonon system in sD$_2$ by neutron
time-of-flight measurements at the IN4 (ILL Grenoble) and at the
TOFTOF (FRM II, Munich). These measurements are described in
detail in another paper \cite{Frei-2}.

\begin{figure}
\includegraphics[width=0.5\textwidth]{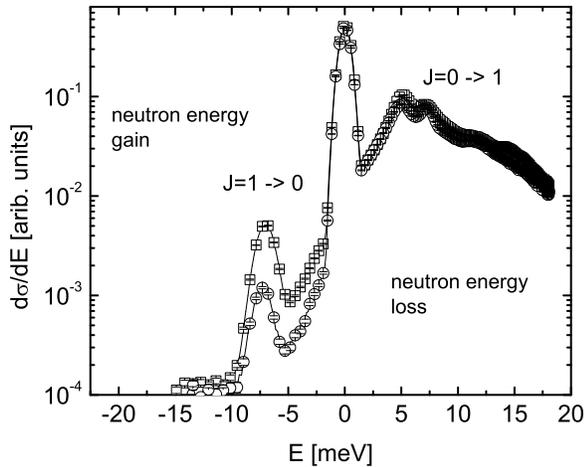}
\caption{An example of a dynamical neutron cross section of solid
D$_2$ at $T=7$ K. Comparison of two ortho concentrations
$c_o$=66.7\%
 ($\square$) and $c_o$=98\% ($\bigcirc$). Data from TOFTOF
measurements at the FRM II. Initial energy of the thermal neutrons
 is E$_0$=20.4~meV.} \label{fig.1}
\end{figure}

Fig. \ref{fig.1} shows two examples of our neutron scattering data
(angular averaged) for two different concentrations c$_o$ of
ortho-D$_2$ molecules. On the side of energy gain of the neutrons
(positive E) we can identify the rotational transition $J=1
\mapsto 0$. The intensity of this peak decreases with reduced
para-D$_2$ molecule concentration $c_p=1-c_o$. A similar
rotational transition (the $J=0 \mapsto 1$ transition) can also be
observed on the neutron energy loss side. Furthermore it is
remarkable that the elastic cross section ($E=$0~meV) of sD$_2$ is
quite large. Our data indicates that the elastic cross section
makes up approximately 50\% of the total cross section at thermal
neutron energy. Another interesting feature is the neutron
scattering intensity on the neutron energy gain side close to the
elastic peak ($E$=0~meV). An increase of the concentration of the
para D$_2$ molecules leads to a larger neutron up-scattering close
to the elastic line. The origin of this cross section is very
likely induced by phonons, which are interlinked with the
rotational transitions $J=1 \mapsto 0$ of the para molecules. This
up-scattering has serious implications to the up-scattering of UCN
in the sD$_2$ converter material, and reduces the achievable
density of UCN in sD$_2$. A detailed analysis of these effects
will presented in a forthcoming paper.

\begin{figure}
\includegraphics[width=0.5\textwidth]{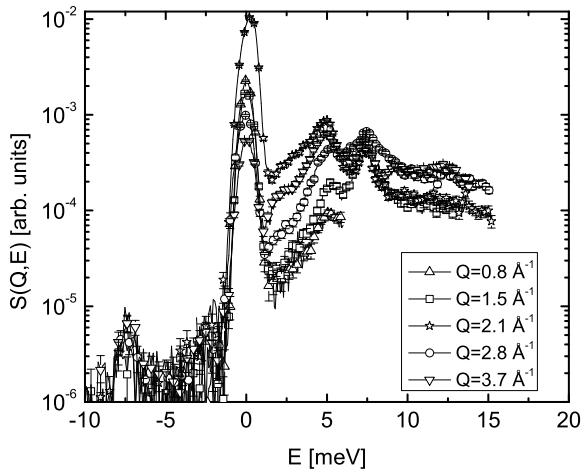} \caption{$S(Q,E)$ of sD$_2$ for
$c_o=95.2\%$ at $T=4$~K for different fixed values of Q. Data from
IN4 measurements. } \label{fig.2}
\end{figure}

\begin{figure}
\includegraphics[width=0.5\textwidth]{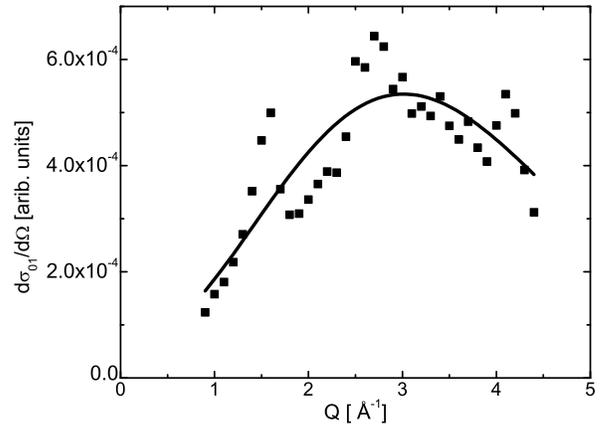} \caption{Cross section $d\sigma_{J=0
\mapsto 1}/d\Omega$ of sD$_2$ for c$_o$=95.2$\%$ o-D$_2$.\
$\blacksquare$ - Data from IN4 measurement. Black line - Fit
result using eq.~\ref{eq.1}.} \label{fig.3}
\end{figure}

In addition the scattering function  $S(Q,E)$ is shown for fixed
values of momentum transfer $Q$ in fig. \ref{fig.2}. The dynamical
scattering function $S(Q,E)$ comprises the phonon branches of the
hcp-sD$_2$ \cite{Nielsen} crystals and the $J=0 \mapsto 1$
rotational transition. The acoustical phonons (transversal and
longitudinal) are clearly visible in $S(Q,E)$ while the optical
phonon branches are not. The rotational transition $J=0 \mapsto 1$
shows a significant Q - dependence and it is purely incoherent.
The position  the $J=0 \mapsto 1$ transition peak
($E_{01}$=7.4~meV) does not change much at different Q-values. The
cross section $d\sigma _{J=0 \mapsto 1}/d\Omega$ for this
transition is shown in fig. \ref{fig.3}. This cross section was
extracted from our data by a integration of S(Q,E) at
$E_{01}$=7.4~ meV with a width of $\Delta E$=~1meV at different
Q-values. The Q-dependence of this cross section was fitted using
the incoherent scattering length of the $J=0 \mapsto 1$ transition
\cite{Frei-2}

\begin{equation}
\label{eq.1}
 \frac{d\sigma _{J=0 \mapsto 1}} {d\Omega}(Q)\sim j_1^2(Qa_s/2)\cdot e^{-\frac{1}{3}
Q^{2}\langle u^{2}\rangle}.
\end{equation}
The parameter $a_s=0.74$~\AA~ is the distance of the deuterons
 within the D$_2$ molecule, while $\langle u^{2} \rangle$ is the mean square
displacement of the D$_2$ molecule in the lattice. The form factor
$j_1$ for the $J=0 \mapsto 1$ transition is described by a
spherical Bessel function of first order. The result for $\langle
u^{2} \rangle \simeq$ 0.245~$\pm$~0.02$ \mathrm{\AA}^2$ is in good
agreement with earlier published results \cite{Nielsen}. The
peak-like deviations in fig.~\ref{fig.3} at certain Q-values are
the result of coherent neutron scattering on phonons.

With the aid of our neutron data we are able to determine the UCN
production cross section by two ways. One way is the determination
of the quasi-particle (phonons and rotational excitations of the
D$_2$ molecules) density of states $G_1(E)$. The other way is the
direct integration of the dynamical neutron cross section in the
kinematical region along the free neutron dispersion parabola.
\\

With the knowledge of the quasi-particle density of states
$G_1(E)$ it is possible to calculate the dynamical neutron cross
section $\frac{d\sigma}{dE_f}$ (averaged over the scattering
angle, thus $Q$). Vice versa it is also possible to extract
$G_1(E)$ from a measured
$\left[\frac{d\sigma}{dE_f}\right]_{data}$ using
\textit{Turchin's} theory \cite{TURCH} for this cross section
applying the incoherent approximation. The absolute normalization
of $\left[\frac{d\sigma}{dE_f}\right]_{data}$ is not needed,
because the extracted $G_1(E)$ has to be normalized to unity
anyway. This approach is only valid within the Born-approximation.

The method of determination of $G_1(E)$ from our data is described
in \cite{Frei-2} in detail. Contributions of higher order
multiphonons to $\frac{d\sigma}{dE_f}$ are incorporated.

\begin{figure}
\includegraphics[width=0.5\textwidth]{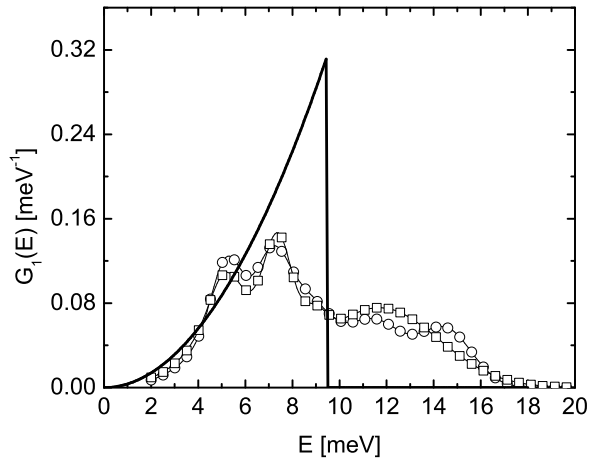} \caption{Comparison of
 the one-quasi-particle density of states of solid D$_2$ for
$c_o=66.7\%$ ($\bigcirc$) and $c_o=98\%$ ($\square$) at $T=7$~K
with a Debye model (line). Data from TOFTOF measurements (energy
resolution: $\Delta E\sim1.24$~ meV). The sD$_2$ poly crystal is
prepared by fast freezing (several minutes) of liquid D$_2$ down
to a temperature of $T<$10~K. $\int_0^\infty{GDOS(E)\cdot dE}=1$}
\label{fig.4}
\end{figure}

The result of our analysis \cite{Frei-2} concerning $G_1(E)$ and a
comparison with a Debye model is shown in fig. \ref{fig.4}. The
characteristics of $G_1(E)$ was already discussed in detail in
\cite{Frei-2} but it is worth to summarizing  these results again,
because they have an important impact on the UCN production cross
section. One major feature is the occurrence of these excitations
above E$\sim$10~meV in the one-quasi-particle density of states
$G_1(E)$. These excitations above E$\sim$10~meV are depending on
the concentration of ortho-D$_2$. The optical phonons in the
region of E~[8-10~meV] are not clear visible in
$G_1(E)$.\\

In the case of UCN production the energy transfer of the
down-scattered neutron $E=E_0-E_f$ is approximately equal to the
initial neutron energy $E_0$ ($E_f=E_U<<E_0$, $E_U$: UCN energy).

\begin{figure}
\includegraphics[width=0.5\textwidth]{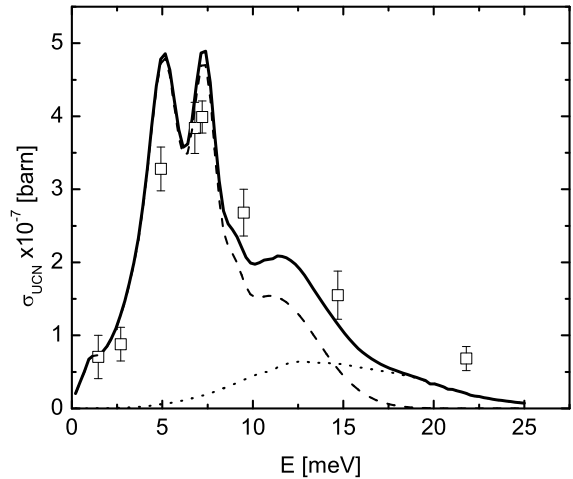} \caption{UCN production cross
section of $c_o=98\%$ solid D$_2$. UCN energy range 0-150 neV
inside the solid D$_2$. Solid line - cross section calculated in
incoherent approximation, using $G_1(E)$ from \cite{Frei-2}.
Dashed line: one-quasi-particle contribution. Dotted line:
two-quasi-particle contribution.
 $\square$ - data from measurements at the PSI \cite{PSI-2}.
} \label{fig.5}
\end{figure}

The total cross section for UCN production can be calculated by

\begin{equation}
\label{eq.2}
 \sigma _{UCN}(E_0)=\int_{0}^{E_{U}^{max}} \frac
{d\sigma (E_0)} {dE} dE_{U}
\end{equation}

and is shown in fig. \ref{fig.5}. It can be compared with recent
published data of the UCN group at the PSI
 \cite{PSI-2} which has studied UCN production at a cold neutron beam ($E_0\sim$1.4~meV to 20~meV).
The agreement of data and our calculation, using our data for
$G_1(E)$ and \textit{Turchin's} incoherent approximation is
surprisingly good. The calculated cross section comprises the
contribution of one-quasi-particle and two-quasi-particle
excitations. Three-quasi-particle excitations do not appear below
E$\sim$14~meV (see fig. 9 in \cite{Frei-2}) and are rather small
in their contribution. The authors of \cite{PSI-2} compared their
results to a calculated cross section based on a multi-phonon
Debye model. They also reported, that using a 'more realistic
model' for the density of states (\textit{Yu et al.}
\cite{Yu,Nielsen}) leads to a considerably worse fit of their
data. Using a Debye model with a cut-off at $k_B\cdot \Theta
_D=$9.5~meV (Debye temperature $\Theta _D$=110~K) only the
contribution of two-phonon excitation have to be included in order
to fit the measured cross section of $\sigma
_{UCN}$=1.55~$10^{-7}$~barn at E=14.7~meV \cite{PSI-2}. This also
applies for the density of states model of \textit{Yu et al.}. Our
own calculated cross section shows a combination of
one-quasi-particle and two-quasi-particle excitations contributing
at these energies. In detail our data indicate (see fig.
\ref{fig.5}) that the UCN production cross section is mainly
determined by one-quasi-particle excitation for energies below
E=15~meV ( see fig. \ref{fig.4}). Even though the
two-quasi-particle contribution can not be neglected in the region
of E~ 5-25~meV.

The application of the incoherent approximation  in the case of
sD$_2$ has certainly to be questioned. The sD$_2$ crystal scatters
neutrons more coherently than incoherently. This fact leads to the
question: Is it possible to get the UCN production cross section
directly from the neutron scattering data?

The easiest way of determining the cross section for UCN
production is the use of the dynamical scattering function
$S(Q,E=\frac {\hbar^2} {2m} Q^2)$ in the $(Q,E)$-phase space along
the free neutron parabola (dispersion curve with $E\simeq
 E_0=\frac {\hbar^2} {2m} k_0^2$ and $Q\simeq k_0$). This approach
equally rely on the Born approximation. The UCN production cross
section can be determined by

\begin{equation}
\label{eq.3}
 \sigma _{UCN}(E_0)=\frac {\sigma _0} {k_0} S(k_0,\frac
{\hbar^2} {2m} k_0^2) \frac {2} {3} k_U^{max} E_U^{max} .
\end{equation}

The term $E\simeq E_0=\frac {\hbar^2} {2m} k^2$ is the energy of
the incoming neutron in the down-scattering process, while
$k_U^{max}$ and $E_U^{max}$ are the upper limits for the UCN
momentum and energy. In order to obtain absolute cross sections
the dynamical scattering function of solid deuterium extracted
from our neutron scattering data has to be calibrated to absolute
values. This can be done by using the rotational transition $J=1
\mapsto 0$ as reference process for calibration. The cross section
for the $J=1 \mapsto 0$ transition was calculated by \textit{
Hamermesh} and \textit{Schwinger} \cite{Schwinger} for ortho- and
para- deuterium molecules (gas). These calculation include also
thermal movements (maxwellian distribution of velocities) of the
deuterium molecules in the gas. The application of these
calculations to the rotational transitions of the D$_2$ molecules,
which are pinned in a crystal at fixed positions may of course be
questionable. However the rotational transitions in sD$_2$ are not
very much hindered by the crystal binding of the crystal
\cite{ROT}. Indeed the molecules are of course not at rest, they
perform motions around their equilibrium positions. These
movements are caused by the zero-point-motions and also by
existing phonon excitations in the crystal. A similar problem was
considered by \textit{Lamb} \cite{Lamb} in the past. He calculated
the capture of neutrons by atoms in crystals. His calculations for
crystals with weak lattice binding delivered the finding, that
atoms in such crystals can be treated as a gas with an effective
temperature, which depends on the Debye temperature of the
crystal. This effective temperature can be higher than the real
temperature of the solid and is just a result of the average
energy of the vibrational degrees of freedom plus the
zero-point-motion. We used this ansatz to calculate the effective
temperature ($T_{eff}\simeq$40~K at a deuterium temperature of
4~K) of the D$_2$ molecules and used it for the calculation of the
rotational transition ($\sigma _{J=1 \mapsto 0}$) of our quasi
free molecules, following the theory of \textit{ Hamermesh} and
\textit{Schwinger}.

The cross section $ \sigma _{J=1 \mapsto 0}$ in pure para
deuterium has a value of $\sigma _{J=1 \mapsto 0}$=0.61 barn for
$E_0$=17.2 meV ( corresponding to the energy of the incoming
neutrons at the IN4 experiment) at a deuterium temperature of
$T$=4~K. The comparison of our neutron scattering data of natural
deuterium ($c_O$=66.7$\%$) and ortho enriched deuterium indicated
a ortho concentration of $c_O$=(95.2~$\pm$~0.3)$\%$.

Scaling with the para concentration of our sample
($c_P=1-c_O=$4.8$\%$) thus leads to a value of $\sigma _{J=1
\mapsto 0}=$0.029~barn for our ortho enriched solid D$_2$ sample.
This value for $\sigma_{J=1 \mapsto 0}$ can be used to calculate a
scaling factor for $S(Q,E)=\kappa \cdot S_{data}(Q,E)$.

By means of this value it is possible to calculate the total
neutron scattering cross section for $E_0$=17.2 meV neutrons:
$\sigma _{tot}(E_0$=17.2 meV$)$=23.4~barn. The assumption is of
course only valid, if the kinematical area of our neutron
scattering experiment covers most of the possible scattering
processes. A good cross check of this value is the calculated
value for the total cross section for solid poly-crystalline
deuterium, using the incoherent approximation of \textit{Turchin}
\cite{TURCH}.

\begin{figure}
\includegraphics[width=0.5\textwidth]{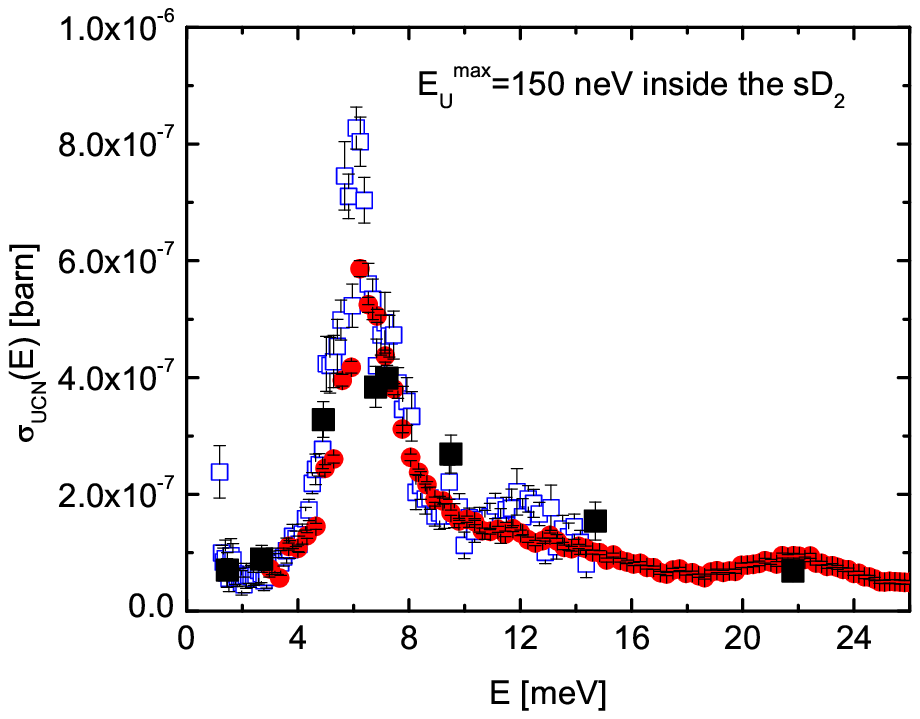} \caption{~}{UCN production cross
section solid D$_2$ of $c_o=95.2\%$. A UCN energy range of 0-150
neV inside the solid D$_2$ is assumed. Cross section determined by
an integration of $S(Q,E)$ along the free dispersion of the
neutron (sample - fast frozen solid deuterium ($T=$4~K)); data
from IN4 measurements. Blue $\square$ - $E_0=$17.2~meV. Red filled
$\bigcirc$ - $E_0=$67~meV, \cite{FN1}.
 $\blacksquare$ - direct UCN production data from measurements
at the PSI \cite{PSI-2}. The PSI data do not show a distinct peak
at $E\sim$~6~meV } \label{fig.6}
\end{figure}

The result of this calculation leads to a value of $\sigma
_{tot}(E_0$=17.2 meV$)$=23.8~barn, which is close to the value of
our calibration. The dynamic response function $S(Q,E)$, resolved
from our data analysis, was not corrected for multiple scattering
effects. Therefore the estimated error of this calibration is
approximately 12$\%$ \cite{MS}. The result of this calibration and
determination of the UCN production cross section as function of
the the energy of the incoming neutrons, and a comparison with
measurements of this cross section \cite{PSI-2} is shown in fig.
 \ref{fig.6}. The agreement of the absolute values of the measured cross
section with the resolved values form our $S(Q,E)$ data is
reasonable. The comparison of the calculated UCN production cross
sections, extracted from the incoherent approximation and parabola
method shows (see fig. \ref{fig.5} and fig. \ref{fig.6}) a
discrepancy in the region of $E\sim$ 6 meV. The cross section,
determined by the parabola method exhibits a pronounced maximum in
the region of $E\sim$ 6 meV as compared to the incoherent
approximation result.

\begin{figure}
\includegraphics[width=0.5\textwidth]{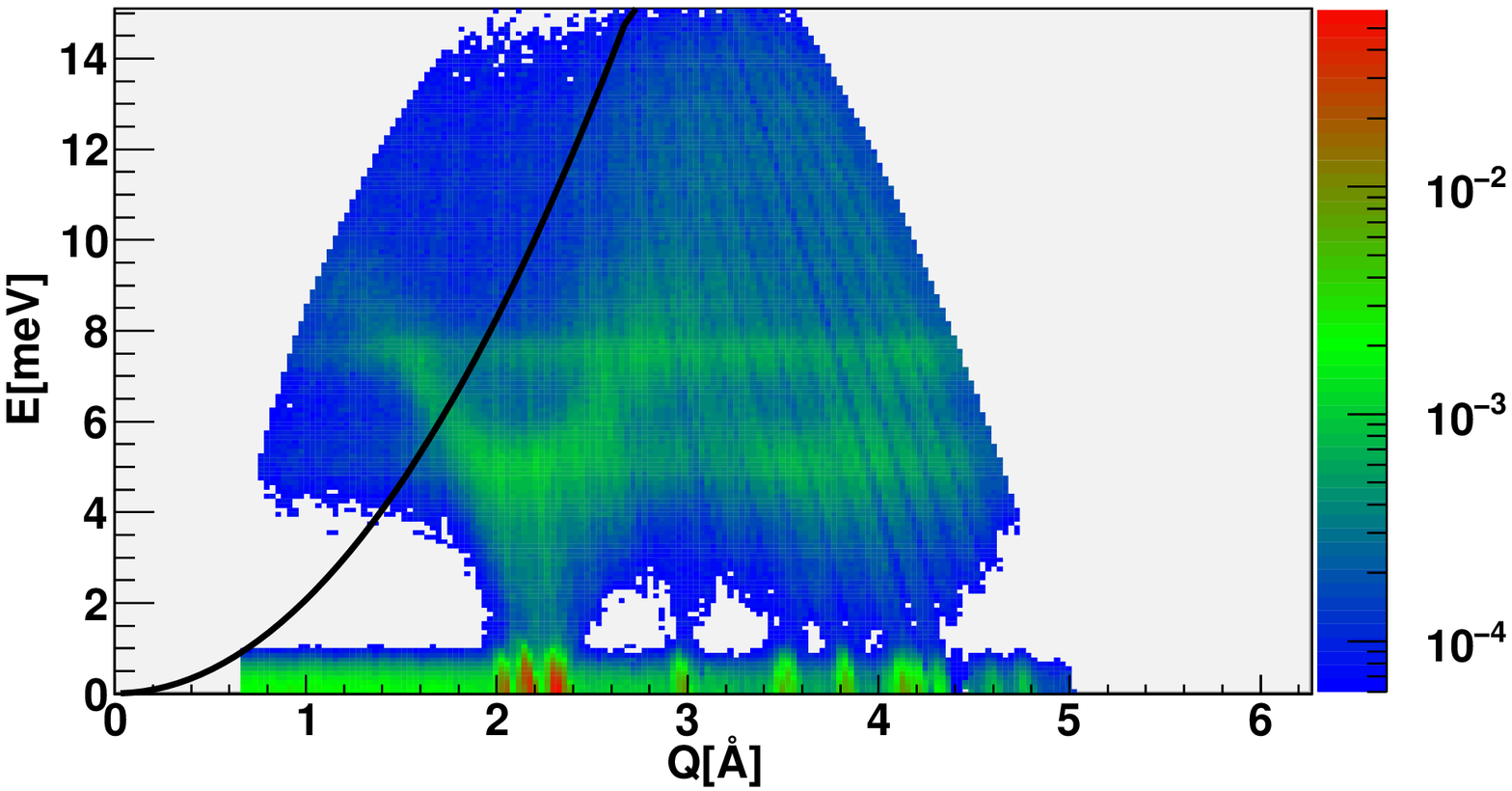} \caption{~}{$S(Q,E)$ of sD$_2$ for
$c_o=95.2\%$ at $T=4$~K. Data from IN4 measurements.
 Black parabola - dispersion of the free neutron.
} \label{fig.7}
\end{figure}

In fig. \ref{fig.7} the dynamical scattering function $S(Q,E)$
(neutron energy loss side) of fast frozen solid ortho-deuterium is
shown. The black line corresponds to the dispersion of the free
neutron.

The parabola of the free neutron crosses  the acoustical phonon
dispersion curve at $E\sim$ 6 meV(see fig. \ref{fig.7}). At this
point, the UCN production cross section is predominantly
determined by coherent scattering. This can explain a deviation
from the production cross section in incoherent approximation.
Nevertheless the general agreement of the  incoherent
approximation with the PSI data is remarkable (see fig.
\ref{fig.5})

The effective UCN production rate P (UCN cm$^{-3}$ s$^{-1}$) from
a neutron beam is determined by integrating the product of the UCN
production cross section $d\sigma/dE$ and the spectral flux
$d\Phi/dE$ (Maxwell spectrum with effective neutron temperature
$T_n$) of the incoming neutrons over the finite UCN energy range
($E_{U}^{max}$ maximum allowable UCN energy) and over the energy
spectrum of the incoming neutron, still assuming $E=E_0$.

\begin{equation}
\label{eq.4}
 P(T_n)=N_{D_2} \cdot \int_{0}^{E
_{U}^{max}}\int_{0}^{E^{max}} \frac {d\Phi (T_n)}{dE_0} \cdot
\frac {d\sigma} {dE_0} dE_{U} dE_0.
\end{equation}

\begin{figure}
\includegraphics[width=0.5\textwidth]{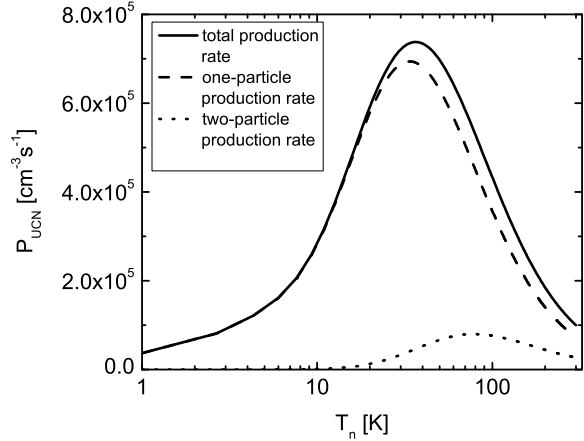} \caption{Calculated UCN production
rate of $c_o=98\%$ solid D$_2$ for different Maxwellian neutron
spectra with effective neutron temperature $T_n$. UCN energy range
- 0-150 neV inside the solid D$_2$. Neutron capture flux $\Phi
_C=1\cdot 10^{14}$~cm$^{-2}$~s$^{-1}$. Solid line - total
production rate (one- and two particle excitation). Dashed line -
one-particle production rate. Doted line - two-particle production
rate.} \label{fig.8}
\end{figure}

In fig. \ref{fig.8} the result for the UCN production rate in
solid ortho deuterium, exposed to Maxwellian shaped neutron flux
for different effective neutron temperatures is shown, using
eq.~\ref{eq.4}.

The main conclusion from these results is the new understanding of
possible higher energetic loss channels (one-quasi-particle and
two-quasi-particle excitations) in solid deuterium for the down
scattering of thermal or cold neutrons in the conversion process
to UCN. The best value for the effective neutron temperature is in
the region of $T_n\sim $40~K, which is larger than the value
($T_n\sim $30~K) reported by \textit{Yu et al.} \cite{Yu} in an
earlier publication. A similar result was obtained by
\textit{Serebrov et al.} \cite{Ser-3} using a Debye-model for
sD$_2$ in a theoretical calculation, based on the incoherent
approximation. The agreement of our result with the results of the
calculation applying a Debye-model is not a surprise, because they
include mulit-phonon excitations in their calculations, while
\textit{Yu et al.} considered only one-phonon excitations.

In summary, new neutron scattering data of solid deuterium leads
to a better understanding of UCN production in this converter
material. The new results for the density of states in sD$_2$ and
the results for the UCN production cross section, extracted
directly from the dynamical scattering function $S(Q,E)$ predict a
significant UCN production cross section for incoming neutrons
with energies higher than $E_0>$~10 meV. This observation was also
confirmed \cite{PSI-2} by direct measurements of the UCN
production cross section. An optimized sD$_2$ UCN source should be
exposed to a cold neutron flux with an effective neutron
temperature of
$T_n\simeq$40~K.\\

\acknowledgments This work was supported by the cluster of
excellence "Origin and Structure of the Universe" (Exc 153) and by
the Maier-Leibnitz-Laboratorium (MLL) of the Ludwig-\\Maximilians-
Universit\"at (LMU) and the Technische Universit\"at M\"unchen
(TUM). We thank T. Deuschle, E. Karrer-M\"uller, S. Materne and H.
Ruhland for their help during the experiments.

\newpage 
\bibliography{UCNP-3}

\begin{thebibliography}{3}
\bibitem[{\citenamefont{FN1}(2009)}]{FN1}
 \bibinfo{author}{\bibfnamefont{The calibration of $S(Q,E)$ at this energy was done by using the 
 calculated total cross section $\sigma _{tot}(E_0=$67~meV$)$~ in incoherent approximation}}.
\bibitem[{\citenamefont{MS}(2009)}]{MS}
 \bibinfo{author}{\bibfnamefont{The treatment of multiple scattering effects in our solid deuterium samples are discussed in detail
 in ref. \cite{Frei-2}}}.
\end{thebibliography}

\end{document}